\documentclass[twocolumn,prl,showpacs,superscriptaddress]{revtex4}
\usepackage{graphicx}
\usepackage{amsmath}
\usepackage{bm}
\usepackage{float}
\usepackage{color}

\begin{document}

\title{Spin squeezing: \\
transforming one-axis-twisting into two-axis-twisting}
\author{Y. C. Liu$^{\dag}$}
\affiliation{Department of Physics, Beijing Jiaotong University,
Beijing 100044, P. R. China}
\author{Z. F. Xu}
\affiliation{State Key Laboratory of Low Dimensional Quantum Physics£¬
Department of Physics, Tsinghua University, Beijing 100084, P. R. China}
\author{G. R. Jin}
\affiliation{Department of Physics, Beijing Jiaotong University,
Beijing 100044, P. R. China}
\author{L. You}
\affiliation{State Key Laboratory of Low Dimensional Quantum Physics£¬
Department of Physics, Tsinghua University, Beijing 100084, P. R. China}

\date{\today}

\begin{abstract}
Squeezed spin states possess unique quantum correlation or
entanglement that are of significant promises for advancing quantum
information processing and quantum metrology.
In recent back to back publications [C. Gross \textit{et al, Nature}
\textbf{464}, 1165 (2010) and Max F. Riedel \textit{et al,
Nature} \textbf{464}, 1170 (2010)], reduced spin fluctuations
are observed leading to spin squeezing
at $-8.2$dB and $-2.5$dB respectively in two-component
atomic condensates exhibiting one-axis-twisting
interactions (OAT). The noise reduction limit for the OAT interaction
scales as $\propto 1/{N^{2/3}}$, which for a condensate with
$N\sim 10^3$ atoms, is about 100 times below standard quantum limit.
We present a scheme using repeated Rabi pulses capable of transforming
the OAT spin squeezing into the two-axis-twisting type,
leading to Heisenberg limited noise reduction $\propto 1/N$,
or an extra 10-fold improvement for $N\sim 10^3$.
\end{abstract}

\maketitle

Squeezed spin states (SSS) \cite{ss1,ss2} are
entangled quantum states of a collection of spins in which the
correlations among individual spins reduce quantum
uncertainty of a particular spin component
below the classical limit for uncorrelated particles \cite{ss1}.
Research in SSS is a topical area due to its significant applications in
high-precision measurements
\cite{ss2,nat1,nat2,clock,nist,nist2,loyd,blatt} and in quantum
information science \cite{sor,Sorensen,xgwang,lewen,smerzi}.
Squeezed spin states were first introduced
by Kitagawa and Ueda, who considered two ways to produce them.
The simplest to implement uses a ``one-axis twisting" (OAT) Hamiltonian,
but the state it produces does not have ideal squeezing properties.
A more complex approach uses a
``two-axis twisting" (TAT) Hamiltonian and produces an improved state.
Other mechanisms for producing SSS have also been investigated,
especially those based on atom-photon interactions \cite{Hald,Takahashi2}
and quantum non-demolition measurements \cite{Kuzmich,Jessen,Polzik,Takahashi}.


Atomic Bose-Einstein condensates are promising systems for observing
spin squeezing. Assuming fixed spatial modes, condensed atoms
are described by a collection of pseudo-spin $1/2$ atoms,
with spin up ($\left|\uparrow\right\rangle $)
and down ($\left|\downarrow \right\rangle $) denoting
the two internal states or spatial modes \cite{han,num,Sorensen,law,Jin}.
The two recent experiments \cite{nat1,nat2} raise significant hope for
reaching the theoretical limit of spin squeezing $\propto
1/{N^{2/3}}$ with $N$ the total number of atoms
for the OAT model \cite{ss1}. Both experiments utilize
two internal hyperfine states of condensed atoms, with the OAT
interaction cleverly constructed from binary atomic collisions,
possibly accompanied by systematic and fundamental imperfections
not confined to the two state/mode approximation. They can be
further degraded by atomic decoherence and dissipation
\cite{nat1,nat2}.

This Letter describes a readily implementable idea for improved
spin squeezing in the two experiments.
Given the reported OAT model parameters \cite{nat1,nat2},
we propose a coherent control scheme capable of transforming the OAT into
the effective TAT spin squeezing,
leading to a Heisenberg limited noise reduction $\propto 1/N$,
or a further 10 fold improvement for a condensate with $\sim 10^3$ atoms.

With a coherent coupling, the two-component condensate
is described by the following Hamiltonian
\begin{equation}
H=\chi J_{z}^{2}+\Omega(t) J_{y},
\label{1ax}
\end{equation}%
as in \cite{nat1,nat2}. The collective spin $\vec{%
J}$ is defined according to ${J}_{v}=\sum_{k}\hat{\sigma}_{v}^{(k)}/2$ in
terms of the Pauli matrices $\hat{\sigma}_{v}^{(k)}$ ($v=x$, $y$, $z$)
for the pseudo-spin of the $k$-th atom. The first term on
the right-hand side of Eq. (\ref{1ax}) is the nonlinear interaction
responsible for the OAT-type spin
squeezing, with $\chi $ the atomic interaction parameter.
The Rabi frequency $\Omega (t)=\Omega _{0}f(t)$ results from
near-resonant two-photon (microwave+rf) coupling between the hyperfine
states \cite{nat1,nat2}. Its maximum amplitude $\Omega_{0}$ and the temporal
envelop $f(t)$ can be controlled experimentally \cite{nat1,nat2}.

\begin{figure}[tbp]
\begin{center}
\includegraphics[width=\columnwidth,angle=0]{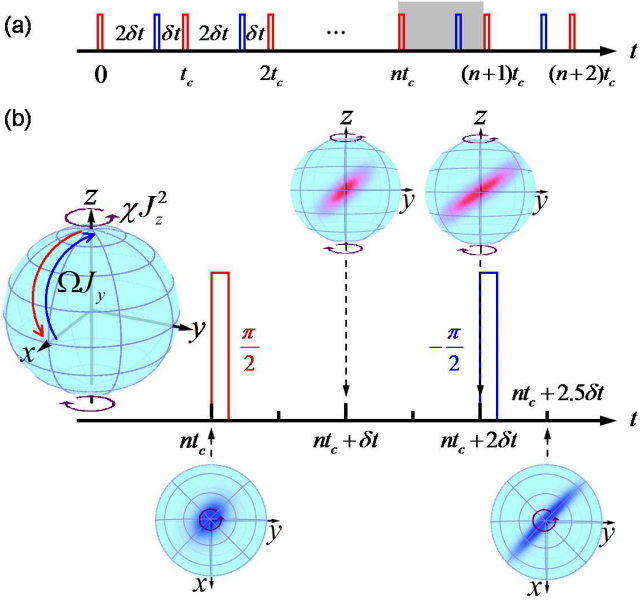}
\end{center}
\caption{An illustration of the proposed pulse sequence
with pulse envelop $\Omega /\Omega_{0}$ vs. time $t$ in arbitrary units,
and the associated spin distributions on Bloch spheres.
(a) The multi-pulse sequence is cyclic, with
one period between $nt_{c}$ and $(n+1)t_{c}$ (shaded) and zoomed in (b).
Each period contains two $\pi/2$ pulses.
A red pulse rotates the spin from along the positive $z$-axis to the
$x$-axis, and a blue one coherently phased to achieve
the opposite, is a $-\pi/2$ pulse, as
shown on the large Bloch sphere of (b).
The four small Bloch spheres are located at
their corresponding times. The spin distributions for the upper
two are centered on the $x$-axis, while the lower two are
centered on the $z$-axis, due to the applications of red and blue
$\pi/2$ pulses. The twistings from the OAT interaction (\ref{1ax})
are illustrated with counter-rotating circular arrows
pierced by the $x$- or $z$-axis of all Bloch spheres.}
\label{figl}
\end{figure}

Our idea for transforming the OAT (\ref{1ax}) into the
TAT makes use of multiple $\pi /2$ pulses affected with the coupling
term $\Omega J_{y}$. In the Rabi limit, $|\Omega |\gg N|\chi |$,
nonlinear interaction can be neglected
while the collective spin undergoes driven Rabi oscillation. A pulse
with an area of $\pi /2$ corresponds to $\int_{-\infty }^{\infty }\Omega
(t)dt=\pi /2$, which gives the transformation
\begin{equation}
R_{-\pi /2}e^{iJ_{z}^{2}}R_{\pi /2}=e^{iJ_{x}^{2}},
\end{equation}%
where $R_{\theta }=e^{-i\theta J_{y}}$, rotating the spin
around the y-axis by an angle $\theta$.
The proposed pulse sequence is periodic as shown in Fig.~\ref{figl}(a).
Each period [Fig.~\ref{figl}(b)] is made up of the following:
a $\pi /2$ rotation about $y$-axis (red pulse), a free evolution for $2\delta t$,
a second $-\pi /2$ rotation about $y$-axis (blue pulse),
and a second free evolution for $\delta t$.
The period is $t_{c}\approx 3\delta t$,
neglecting the time needed for affecting the two $\pm \pi /2$ pulses.

The time evolution operator at time $t=nt_{c}$ ($n=1,2,3,\cdots$), \textit{i.e.},
after $n$ periods, is given by $U^{n}$, where
\begin{equation}
U=e^{-i\tau J_{z}^{2}}R_{-\pi /2}e^{-2i\tau J_{z}^{2}}R_{\pi /2}=e^{-i\tau
J_{z}^{2}}e^{-2i\tau J_{x}^{2}},
\label{U}
\end{equation}%
is for a single period and $\tau = \chi\delta t$.
Using Baker-Campbell-Hausdorff formula, we find
\begin{equation*}
U=e^{-i\tau (2J_{x}^{2}+J_{z}^{2})}\exp [i\tau ^{2}\{J_{x},\{J_{y},J_{z}\}\}+%
\mathcal{O}(\tau ^{2})],
\end{equation*}%
where $\{A,B\}=AB+BA$ denotes anti-commutator for operators $A$ and $B$.
Expanding for small $\tau$,
we arrive at $\exp [i\tau ^{2}\{J_{x},\{J_{y},J_{z}\}\}+\mathcal{O}(\tau ^{2})]\simeq 1$
after neglecting higher order terms.
Hence we end up with $U\simeq e^{-i\tau (2J_{x}^{2}+J_{z}^{2})}$,
and the time evolution operator $U^{n}\simeq e^{-in\tau
(2J_{x}^{2}+J_{z}^{2})}=e^{-i\chi (2J_{x}^{2}+J_{z}^{2})t/3}$,
equivalent to that given by an effective Hamiltonian
$
H_{\mathrm{eff}}={\chi }(2J_{x}^{2}+J_{z}^{2})/3={\chi }%
(J_{x}^{2}-J_{y}^{2}+\vec{J}^{2})/3.
$
After dropping a constant $\vec{J}^{2}=j(j+1)$,
$H_{\mathrm{eff}}$ reduces to that of the TAT.
This shows the application of our proposed $\pi /2$ pulse sequence effectively
transforms the OAT models of the recent experiments \cite{nat1,nat2} into
the TAT models, provided $\tau \ll ({2N)}^{-1}$ \cite{unpub}.
For a fixed $\chi$, the time of optimal spin squeezing
from the TAT $\propto {\chi }(J_{x}^{2}-J_{y}^{2})$
is about $1/3$ of that from the OAT $\propto{\chi }J_{z}^{2}$ with $N=1250$,
i.e., squeezing occurs around $3$ times faster for the TAT.
Thus, despite of the three times reduction in the
effective strength (${\chi }/3$) of the transformed
Hamiltonian $H_{\mathrm{eff}}$, the time for observing
the optimal SS remains almost the same. Consequently,
we expect degradation from atomic losses will be similar to
the case of OAT \cite{yun}.

The validity for our idea of transforming the OAT into the TAT can be directly
checked through comparing the effective dynamics from $H_{\mathrm{eff}}$
with the actual dynamics due to (\ref{1ax}) accompanied with the sequence of pulses.
For this purpose, we expand the state vector time evolved from
the initial state $|j,j\rangle $,
\begin{equation}
\left\vert \Psi (t )\right\rangle =e^{-i\chi t
(J_{x}^{2}-J_{y}^{2})/3}|j,j\rangle =\sum\nolimits_{m}^{\prime }c_{m}(t
)\left\vert j,m\right\rangle ,
\label{seo}
\end{equation}%
into eigenstate of $J_{z}$ with $j=N/2$.
The primed summation implies $m=-j,-j+2,\cdots ,j$,
or $m=-j+1,-j+3,\cdots ,j$, for even or odd $N$ respectively \cite{ss1}.

First we consider the case of small $N$, e.g., $N=2$.
The quantum dynamics
of the effective TAT Hamiltonian $H_{\mathrm{eff}}=$ ${\chi }(J_{x}^{2}-J_{y}^{2})/{3}$
give rise to time evolved probability amplitudes $c_{-1}(t)=-i\sin (\chi t/3)$, $%
c_{0}(t)=0$, and $c_{1}(t)=\cos (\chi t/3)$. For each period $%
t_{c}=3\delta t$ of the exact dynamics
given by Eq. (\ref{U}), the probability amplitudes become
$c_{-1}(t)=e^{-i\chi (t+nt_c)}/2$, $c_{0}(t)=%
e^{-i\chi \delta t}/{\sqrt{2}}$, $c_{1}(t)=e^{-i\chi (t+nt_c)}/2$ for $%
(n-1)t_{c}<t\leq (n-1)t_{c}+2\delta t$, and $c_{-1}(t)=-ie^{-i\chi
(t-nt_{c}/3)}\sin (\chi nt_{c}/3)$, $c_{0}(t)=0$, $c_{1}(t)=e^{-i\chi
(t-nt_{c}/3)}\cos (\chi nt_{c}/3)$ for $nt_{c}-\delta t<t\leq nt_{c}$.
At $t=nt_{c}$, \textit{i.e.}, after $n$ periods, they
become $c_{-1}(t)=-ie^{-i2\chi t/3}\sin (\chi t/3)$, $c_{0}(t)=0$,
and $c_{1}(t)=e^{-i2\chi t/3}\cos (\chi t/3)$, exactly the same as
that from the effective TAT dynamics, apart from an overall phase
for the $\vec J^2$ term.

\begin{figure}[tbph]
\centerline{
\includegraphics[width=\columnwidth,angle=0]{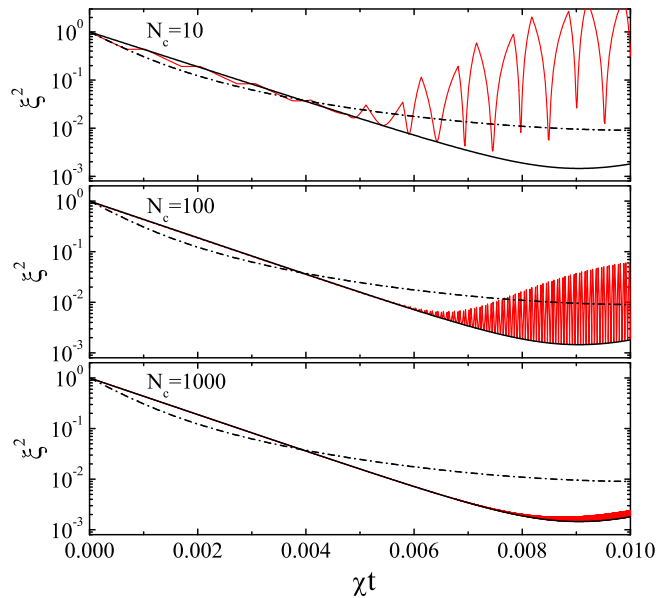}}
\caption{Spin squeezing parameter at different pulse numbers
$N_c=10$ (upper), $100$ (middle), and $1000$ (lower) compared with
the effective TAT $H_{\mathrm{eff}}$ dynamics in black solid lines
with the actual ones (\ref{1ax}) using the proposed sequence of pulses
in red solid lines.
The black dot-dashed lines denote OAT results,
all for $N=1250$ atoms.}
\label{fig2}
\end{figure}

Next we consider larger $N$ numerically,
e.g., $N=1250$ atoms and for different number of pulses $N_c$
during the time of optimal squeezing $\chi t\approx 3\ln(4N)/(2N)$ \cite{unpub}.
The TAT Hamiltonian $H_{\mathrm{eff}}$ can generate
optimal spin squeezing when applied to the initial state
$|j,j\rangle =\left|\uparrow \right\rangle ^{\otimes N}$ \cite{ss1}.
The state expansion Eq. (\ref{seo}) involves basis $|j,m\rangle$
with even or odd $m$
because of the conserved parity $\exp (i\pi J_{z})$.
As a result, the mean spin is always along the $z$-axis,
i.e., $\langle \vec{J}\rangle =(0,0,\langle J_{z}\rangle)$.
For any spin component normal to $\langle \vec{J}\rangle $, $J_{\gamma
}=J_{x}\cos \gamma +J_{y}\sin \gamma $ with an arbitrary angle $\gamma $,
its variance is found to be
\begin{equation}
(\Delta J_{\gamma })^{2}=\frac{1}{2}[\mathcal{C}+\mathcal{A}\cos (2\gamma )+%
\mathcal{B}\sin (2\gamma )],
\end{equation}%
where $\mathcal{A}=\langle J_{x}^{2}-J_{y}^{2}\rangle =3\chi ^{-1}\langle
H_{\rm eff}\rangle $, $\mathcal{B}=\langle J_{x}J_{y}+J_{y}J_{x}\rangle =$Im$\langle
J_{+}^{2}\rangle $, and $\mathcal{C}=\langle J_{x}^{2}+J_{y}^{2}\rangle
=j(j+1)-\langle J_{z}^{2}\rangle $. The optimal squeezing angle $\gamma _{%
\mathrm{op}}$ is obtained from minimizing $(\Delta J_{\gamma })^{2}$ with
respect to $\gamma $, yielding $\tan (2\gamma _{\mathrm{op}})=\mathcal{B}/%
\mathcal{A}$. As the coefficient $\mathcal{A}=$Re$\langle J_{+}^{2}\rangle
=0 $, we have $\gamma _{\mathrm{op}}=\pi /4$ or $3\pi /4$, which implies
that squeezing and anti-squeezing occurs for $J_{x}\pm J_{y}$, \textit{i.e.},
along the angle bisector of $x$- and $y$- axis, with the reduced
variance $V_{-}=(\mathcal{C}-|\mathcal{B}|)/2$. Alternatively, the increased
variance $V_{+}=(\mathcal{C}+|\mathcal{B}|)/2$ is associated with
anti-squeezing.

The reported degree of squeezing is measured by
spin squeezing parameter $\zeta_{\rm S}^2=NV_-/\langle \vec J\rangle^2$
from Ramsey spectroscopy \cite{ss2},
which differs slightly from $\xi^2=2V_-/j$.
The latter form is used to graph spin fluctuations in
the relevant figures of Refs. \cite{nat1,nat2}, which is
independent of angular momentum coordinate system
or specific measurement scheme.
For the coherent spin state $|j,j\rangle $, the variances $(\Delta J_{\gamma
})^{2}=j/2$ and $\xi ^{2}=1$. Time evolution from
$H_{\mathrm{eff}}$, develops quantum correlation, which transforms
$|j,j\rangle$ into a SSS with $\xi^{2}<1$. As shown in Fig.
\ref{fig2}, with the increase of pulse number $N_c$, the actual
OAT dynamics approach and eventually settle down to the effective
dynamics of $H_{\rm eff}$. The minimum number
of periods required before reaching the optimal squeezing point is
found analytically and checked numerically to satisfy $N_c\gg \ln (4N)$ \cite{unpub}
in order to generate the
effective dynamics of the TAT. For $N=1250$, $\ln (4N)\approx 8.5$, thus
it is reasonable to expect a near perfect agreement at $N_c=1000$.
In practice, without special control techniques,
errors from repeated pulses can build up, limiting
the performance of our proposal.

\begin{figure}[tbph]
\centerline{
\includegraphics[width=\columnwidth,angle=0]{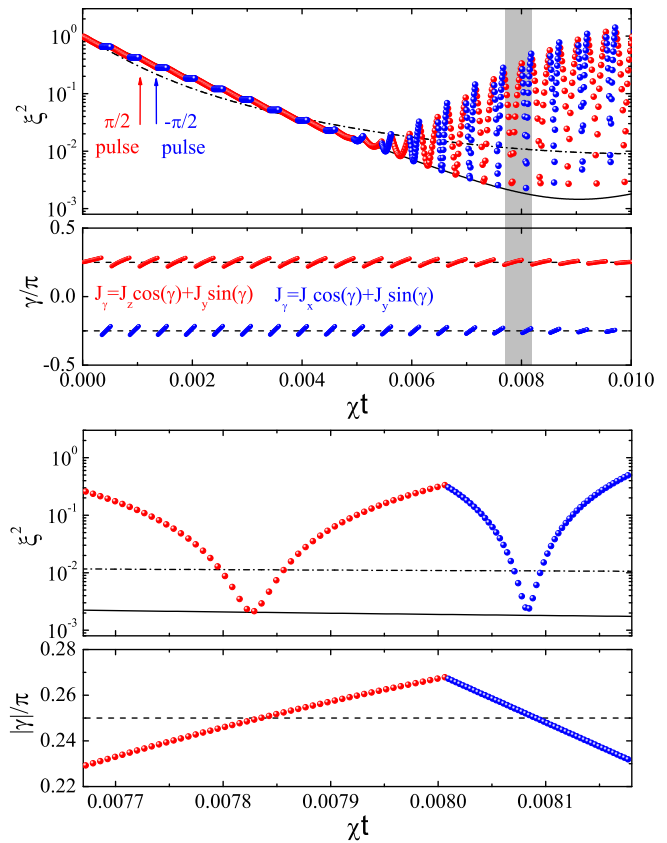}}
\caption{Top-higher panel: Spin squeezing parameter
calculated from the actual dynamics of (\ref{1ax}) for $N_c=20$ pulses
shown by red and blue dots respectively for time windows after the first
(red) and second (blue) pulses of each cycle.
For comparison, the black solid (dot-dashed)
line denotes results from the associated TAT (OAT) model.
The top-lower panel shows the corresponding optimal
squeezing measurement angle.
The dichotomy of two narrow distributions around $\pm\pi/4$ is due to
the repeated $\pi/2$ pulses rotating between the $x$- and $z$-axis.
For the TAT around the $z$- or $x$-axis, this angle is simply a flat distribution at $\pm\pi/4$,
as shown by black dashed lines.
At the reduced number of cycles $N_c=20$, the TAT is not yet
fully effective, which is the cause for small slopped distributions through $\pm\pi/4$.
The slops eventually reduce to zero with the increase of $N_c$.
The bottom panels display in detail
the shaded region where the slops are now displayed in absolute values, all for $N=1250$ atoms.}
\label{fig3}
\end{figure}

Fortunately, however, we find that even for significantly fewer
number of pulses, the TAT-type spin squeezing limit can still be
reached at selected times before arriving at the optimal squeezing.
This is presented in the upper panel of Fig. \ref{fig3}, where spin squeezing
obtained from the actual dynamics (\ref{1ax}) are shown by
red and blue dots respectively for times after the first (red)
and second (blue) pulses of each period. One can see clearly
that in segments of red and blue dots, the oscillating
squeezing parameter kisses the TAT spin squeezing represented by the
black solid curve repeatedly.
Furthermore, from a technical point of view, the measurement
of spin squeezing can be carried out more straightforwardly
as the angle for optimal squeezing $\gamma_{\rm op}$
is essentially fixed, corresponds to
a nearly flat distribution in the immediate temporal neighborhood
as shown in the lower panel of Fig. \ref{fig3}.

The dynamic behavior of spin squeezing from the Hamiltonian
(\ref{1ax}) has been studied before \cite{law,Jin}, including the idea
of turning off the OAT interaction \cite{jaksch}, and the projection
into a desired form \cite{Lukin} as in NMR dynamic decoupling.
Our idea of multiple pulses as presented in this work
is straightforward. It comes with a clear physical picture,
and makes use of coherent control techniques
\cite{UDD1,UDD3}. With suitable generalization, our work
suggests that for systems of identical spin-1/2 atoms, any forms of
binary interactions, e.g., the famous Lipkin-Meshkov-Glick
model \cite{scott}, can be transformed into the effective TAT-type
provided their original spin-spin interaction is NOT SU(2) symmetric,
or NOT proportional to $\vec J^2=J_x^2+J_y^2+J_z^2$.
We can further adopt techniques like spin echo
using additional $\pi$-pulses middle way between all
neighboring $\pm\pi/2$-pulses, to effectively suppress
both spin-spin relaxation or other inhomogeneous effects.

Before conclusion, we note that in all our simulations for
the actual dynamics of (\ref{1ax}), we have assumed the $\pm\pi/2$ control
pulses are executed quickly, during which the nonlinear twisting interaction
is negligible. The results presented above in Figs. \ref{fig2} and \ref{fig3}
are found to remain unchanged irrespective of whether square shaped
or Gaussian shaped pulses are used as long as they give the correct
pulse area of $\pi/2$ and are short enough so that spin squeezing
during the pulses can be neglected. When the phase accumulated
from spin squeezing interactions are included, they generally lead to
a gradual degradation of the achievable spin squeezing.
Based on numerical simulations over a broad range of parameters \cite{ss1,You1},
we conclude for large $N$ ($>100$) the time for
optimal squeezing $\chi t_{\rm op} \simeq 1.58\ln (N)/N$ \cite{unpub},
consistent with the earlier result \cite{Andre}.

In conclusion, we present an idea based on coherent control theory.
Adopting repeated Rabi pulses we transform the OAT spin squeezing
observed in the two recent experiments \cite{nat1,nat2} into
stronger and more effective TAT spin squeezing.  Both analytical
analysis and numerical simulations are presented that confirm the validity
of our proposal. The Heisenberg limited noise reduction $\propto 1/N$
from our proposal is an additional factor $\propto 1/N^{1/3}$
lower than that reached by the OAT model \cite{ss1}.
This work thus enables improved spin squeezing from $\propto 1/N^{2/3}$ to $\propto 1/N$
despite a factor of $1/3$ reduction in the effective
atomic nonlinear interaction strength. Additionally, the
optimal squeezed direction/angle for the TAT interaction is free from
the swirling associated with the OAT \cite{ss1,nat1,nat2}, allowing
for simpler and cleaner measurements.
Our key idea of controlled dynamics to affect a TAT interaction Hamiltonian
from a OAT form can be further generalized to other systems where
OAT spin squeezing are discussed \cite{Sorensen,add1,add2}.
Our proposal calls for no major complications to the available
experimental setups, except for a more advanced timing sequence
for executing Rabi pulses while keep tracking of their phases.
Thus we fully expect that our idea can be realized within current experiments
and hopefully implemented immediately. The experimental
demonstration of our proposal could significantly push the frontier
of quantum metrology with squeezed spin states into new territory.



This work is supported by the NSFC (Contracts No.~10804007 and
No.~11004116) and the SRFDP (Contract No.~200800041003). L.Y. is
supported by the NKBRSF of China and by the
research program 2010THZO of Tsinghua University..

\end{document}